\documentclass[conference]{IEEEtran}
\usepackage{tikz}
\usepackage{amsmath}
\usepackage{amsthm}
\usepackage{subcaption}
\usepackage{xspace}
\usepackage{xcolor}
\usepackage{subfiles}
\usepackage{multirow}
\usepackage{subfiles}
\usepackage{algorithm,algpseudocode}
\usepackage{cite}
\usepackage{amsmath,amssymb,amsfonts}
\usepackage{subfiles}
\usepackage{graphicx}
\usepackage{textcomp}
\usepackage{xcolor}
\usepackage{xspace}
\usepackage{stfloats}
\usepackage{float}
\usepackage{mathtools}
\usepackage{envmath}
\usepackage{algorithm,algpseudocode}
\usepackage{tabularx}
\usepackage{tabularx}
\usepackage{subcaption}
\usepackage{booktabs}
\usepackage{array}
\usepackage{hyperref}
\usepackage[utf8]{inputenc}

\newif\ifshowcomments
\showcommentstrue

\ifshowcomments
\newcommand{\mynote}[2]{\fbox{\bfseries\sffamily\scriptsize{#1}}
	{\small$\blacktriangleright$\textsf{\emph{#2}}$\blacktriangleleft$}}
\else
\newcommand{\mynote}[2]{}
\fi

\definecolor{cadmiumgreen}{rgb}{0.0, 0.42, 0.24}
\newcommand{\contribution}{c}

\newcolumntype{L}[1]{>{\raggedright\let\newline\\\arraybackslash\hspace{0pt}}m{#1}}
\newcolumntype{C}[1]{>{\centering\let\newline\\\arraybackslash\hspace{0pt}}m{#1}}
\newcolumntype{R}[1]{>{\raggedleft\let\newline\\\arraybackslash\hspace{0pt}}m{#1}}
\DeclareOldFontCommand{\bf}{\normalfont\bfseries}{\mathbf}

\theoremstyle{definition}

\theoremstyle{remark}

\begin{document}
%
\title{An Exploratory Analysis on Users' Contributions in Federated Learning}

\author{
\IEEEauthorblockN{Jiyue Huang$^*$}
\IEEEauthorblockA{Delft University of Technology\\
Netherlands\\
Email: J.Huang-4@tudelft.nl}
\\
\IEEEauthorblockN{Sara Boucchenak}
\IEEEauthorblockA{INSA-Lyon\\
France\\
Email: Sara.Bouchenak@insa-lyon.fr}
\and
\IEEEauthorblockN{Rania Talbi}
\IEEEauthorblockA{INSA-Lyon\\
France\\
Email: rania.talbi@insa-lyon.fr}
\\
\IEEEauthorblockN{Lydia Y. Chen$^*$}
\IEEEauthorblockA{Delft University of Technology\\
Netherlands\\
Email: lydiaychen@ieee.org}
\and
\IEEEauthorblockN{Zilong Zhao}
\IEEEauthorblockA{Delft University of Technology\\
Netherlands\\
Email: Z.Zhao-8@tudelft.nl}
\\
\IEEEauthorblockN{Stefanie Roos$^*$}
\IEEEauthorblockA{Delft University of Technology\\
Netherlands\\
Email: s.roos@tudelft.nl}
}


%

\maketitle

\begin{abstract}

Federated Learning is an emerging distributed collaborative learning paradigm adopted by many of today's applications, e.g., keyboard prediction and object recognition. 
Its core principle is to learn 
from large amount of users data while preserving data privacy by design as collaborative users only need to share the machine learning models and keep data locally. The main challenge for such systems is to provide incentives to users to contribute high-quality models trained from their local data.
In this paper, we aim to answer how well incentives recognize (in)accurate local models from honest and malicious users, and perceive their impacts on the model accuracy of federated learning systems. 
We first present a thorough survey on two contrasting perspectives: incentive mechanisms to measure the contribution of local models by honest users, and malicious users to deliberately degrade the overall model. 
We conduct simulation experiments to empirically demonstrate if existing contribution measurement schemes can disclose low-quality models from malicious users. Our results show there exists a clear tradeoff among measurement schemes in terms of the computational efficiency and effectiveness to distill the impact of malicious participants. We conclude this paper by discussing the research directions to design resilient contribution incentives.

\end{abstract}


\section{Introduction}
\label{sec:Introduction}
The increasing capabilities of ubiquitous sensors and smart devices, whether in terms of computation, storage, or connectivity resources, are driving services from the cloud side to the edge of the networks~\cite{shi2016edge}.
Popular machine learning (ML) services are no exception to this trend.
Another critical reason behind this trend is the privacy concern~\cite{mohassel2017secureml} of user data that is often sensed and collected on edge devices.  Users increasingly ask for on-device learning so as to minimize sharing the data with the cloud. 

\begin{figure}
\setlength{\abovecaptionskip}{-0cm}  
\includegraphics[height=6.5cm]{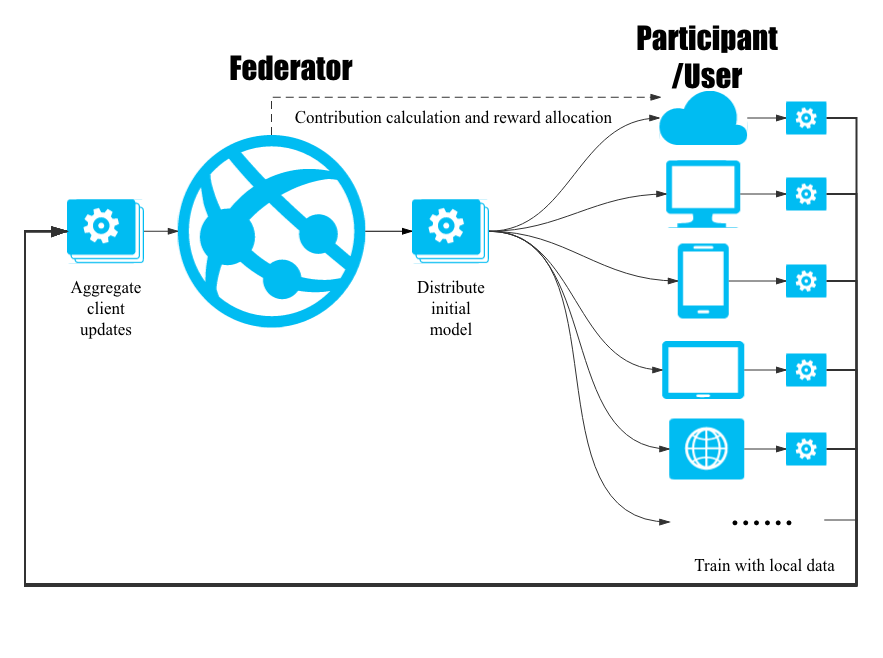}
\caption{An illustration of federated system: federator and multiple users/participants.}
\label{fig:FL}
\end{figure}

Federated Learning (FL)~\cite {yang2019federated} is the emerging paradigm that empower ML-tasks on edge devices in a privacy-preserving manner. 
FL systems enable collaborative training of a machine learning model across distributed users by local model sharing, instead of direct data exchange with the  untrusted service providers. Figure~\ref{fig:FL} illustrates a simplified federated system, where there are multiple users and one federator, the light-weight central server to measure the contribution and provide the rewards\footnote{This is one of the most common  configurations of federated systems~\cite{yang2019federated}}. Users rely on their local data to train a common model and periodically exchange their updates of model parameters with the federator, e.g., the weights of neural networks, until the common model converges. 

As local data never leaves the users' devices in federated learning systems, personalized applications that also benefit from other collaborative users thrive, e.g. text prediction~\cite{hard2018text}, voice recognition~\cite{leroy2019federated}, and self-driving cars~\cite{bagdasaryan2020backdoor}. However, in collaborative systems, it is more a norm than rarity that there exist malicious users who either purposely deteriorate the model quality or take advantage of the system without producing real contributions (free-riders).

In this paper, we study the impact of incentive mechanisms on the model quality of federated learning systems considering two type of participants: i)  honest participants with varying update quality and ii) malicious participants who deliberately send low-quality updates.
We show how incentives mechanisms characterize contributions made by these two types of participants and to survey the-state-of-the-art incentive mechanisms that lead to maximal model accuracy.

The specific contributions of this paper are summarized as follows:
\begin{itemize}
\item We provide an exploratory analysis of  contribution measurement and incentive mechanisms in the presence of honest participants (Section~\ref{sec:Incentives}).
\item We characterize malicious behaviors that has been shown to deteriorate model accuracy (Section~\ref{sec:MaliciouscontribQuality}).
\item We experimentally evaluate existing contribution metrics in the presence of malicious participants  (Section~\ref{sec:contribQuality}). 
\item We provide future research directions to better assess users' contribution and hence handle honest and malicious participants (Section~\ref{sec:researchDirections}) . 
\end{itemize}

\section{Background and Preliminary Notions}
\label{ssec:Overview}



\textbf{\textit{Federated Learning}} is a machine learning setting where multiple participants collaborate in solving a machine learning problem, under the coordination of a central server or service provider called \textbf{\textit{federator}}. Each participant’s raw data is stored locally and is not exchanged or transferred; instead, model updates, e.g., weights of  intended for immediate aggregation are used to achieve the learning objective \cite{Kairouz:2019:corr:advance}. 

The federator plays the role of an orchestrator. It starts the training process by assigning
learning tasks to the participants, initializes the \textbf{\textit{global model}}, and aggregates the \textbf{\textit{updates}} submitted by participants in each training round. These updates can be either neural network weights or gradients in existing studies.

\textbf{\textit{Participants}} (or \textit{Users}), on the other hand, locally own data relevant to these specific training tasks. It is important that participants have sufficient computation capability, data, and network resources to be involved in the training process. They use their local training data to update the global model sent by the Federator to build their own \textit{local models}.

\begin{figure}
\setlength{\abovecaptionskip}{-0cm}   
\includegraphics[height=10.0cm]{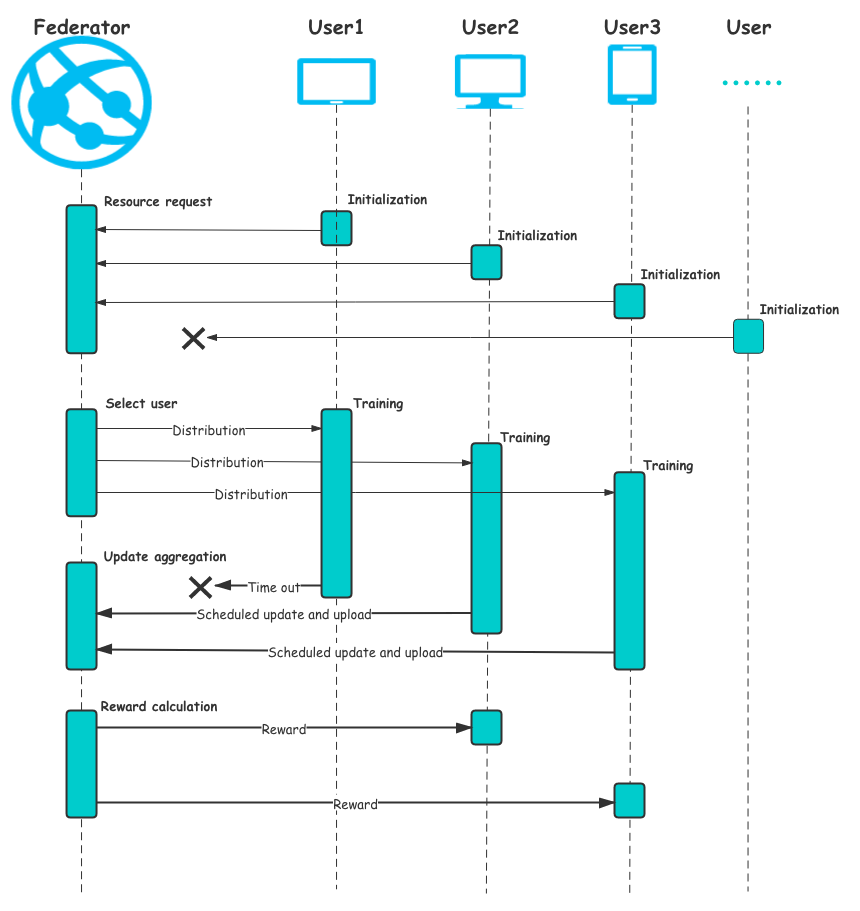}
\caption{Protocol of Federated Learning}
\label{fl_pro}
\end{figure}

Federated learning is an iterative learning procedure composed of five steps that are summarized in Figure~\ref{fl_pro}. These steps are the following:  
\textbf{1. Initialization:} The federator defines a specific machine learning task and initialize the global model. \textbf{2. Participant Selection:} To maximize the model quality and for the sake of fault tolerance, the Federator chooses participants with a good network connection and battery level to take part in the training process at a given round, where one round refers to one iteration of local training and global aggregation along with reward allocation. 
\textbf{3. Local Training:} Selected participants receive the initial model from the federator and train local models using their own data. \textbf{4. Secure Aggregation:} The federator averages the model updates uploaded from partcipants without access to their local data. \textbf{5. Reward Allocation:} The federator distributes rewards to participants based on their own contribution. 
All steps but Initialization are iterated until the global model achieves a desired performance.

In real-world applications, participants can either be honest, whose submitted updates are genuinely trained locally with varying data quality, or malicious. Malicious participants misbehave to gain more profits from the offered service or even aim at deteriorating the whole FL ecosystem. In order to characterize the behavior of both types of participants, we would, however, focus on different peculiarity in response to their presences.
First, for honest users, fair reward distribution mechanism surely encourage users' participation, especially those with high data quality and willing to contribute more computational power.
Designing feasible contribution measurement strategies in federated learning is indispensable but challenging since directly assessing the quality of a user's local data is not possible for the other participants and the federator. Accordingly, there are a number of contribution measurement strategies and corresponding reward systems (see Section~\ref{sec:Incentives}).
In contrast, for malicious users, it is essential to identify the malicious nodes and the type of misbehavior. Based on different classes of attacks, defences need to be designed accordingly. In this paper, we present a thorough classification of both attacks and defences.


\label{sec:contribQuality}

\section{Assessing Contribution for Honest Participants}
\label{sec:Incentives}

For honest Users in federated learning, Federators are supposed to recruit sufficient participants to complete the large-scale tasks with high quality. 
Participants are more willing to provide high-quality data and resources if they receive rewards. The value of the reward should relate to a participant's level of contribution, i.e., participants who contribute more, by some measures, should receive a bigger reward. 
Yet, a major challenge for contribution measurement
of FL systems is data isolation caused by the fact that users keep their raw data secret. Local updates reveal information about their performance indirectly, since parameters of neural networks are deep mapped features and do not carry direct information.
 As a result, FL systems can measure contribution based on updates, without requiring access to the raw data. 

\subsection{Contribution Evaluation Taxonomy}

In this section, we summarize three major taxonomy contribution measurement strategies applied in existing federated learning systems. They are of evidently differences in detecting accuracy and transmission complexity but could be suitable for various of application scenarios.

\subsubsection{\textbf{Test \slash Self-Reported Based Contribution Evaluation}}
The most straight-forward way to measure contribution is to have participants self-report their score, as they have access to their local data and can hence conduct the measurements. 
Theoretically, self-reported contribution is not a measurement
strategy, so we would not discuss it specifically in this paper.

There are multiple ways to define the quality of data in the context of self-reporting. The first one is just the size of the data\cite{feng:2019:ithings:relay}, without knowing their distribution.
So in this paper, the model owner (federator) negotiates with the mobile devices (users) about the size of their training data. In return, each mobile device receives the revenue. 
Alternatively, revenues can depend on the accuracy of the solution to the local sub-problems~\cite{Pandey:twc:2020:OnDevice}

Prior to formally define the measure of users' contributions, we first introduce the notations and assumptions.
We assume there is a linearly decreasing 
valuation function $v(\theta_k)$ (which is negatively related to reward portion) for user $k$ 
depending on the relative accuracy 
$\theta_k$ attained for the local sub-problem. 
The protocol, however, requires a trusted third party to ensure uniform pricing as basis
and leaves it open how such a trusted party would be realized in practice.  

\subsubsection{\textbf{Marginal Loss Based Contribution Evaluation}}
The marginal loss strategies determine the benefit that a participant deserves according to the marginal loss that it brings withdrawing from the alliance. It is widely adopted in Profit Distribution Games~\cite{wang2017profit}, which refers to designing reasonable profit distribution strategies among multiple contributors, such as reward allocation for users in federated learning. 
We  note that computing marginal loss requires a central party, which could be either the federator or a different trusted third party with access to the global model. 
Based on the idea of marginal loss, Richardson et al.\ \cite{Richardson:2019:IJCAIFL:Rewarding} show how a payment structure can be designed to measure contributions of different data owners for linear regression models in a crowd-sourcing scenario as well as assigning incentives.
It determines the influence that data points have on the loss function of the model to calculate the decrease without a specific user owning these data points.  However, the paper merely focuses on linear regression and hence is not of general interest.  Furthermore,  \cite{wang:2019:Bigdata:contribution}  designs a deletion method to measure contribution of horizontal FL,
which means users hold data with same feature space and different ID. 
In contrast, Shapley Value~\cite{Roth:1988:Shapley} has been introduced for vertical FL, referring to users holding data with different feature space and same ID. 
While the Shapley Value can be seen as a marginal loss-based contribution measurement, its main idea relates to game-based incentives, so that we defer to the respective section for a detailed explanation.

\subsubsection{\textbf{Similarity Based Contribution Evaluation}}
Marginal loss-based strategies require the federator or a third party to implement contribution evaluation. However, there are also studies~\cite{Kang:2019:IEEEIoT:contract} that focus on pairwise measurement, i.e., participants evaluate each other. 
In this manner, the system reduces both the trust in and the load on the central party. Having a distributed contribution measurement further enhances robustness to the central failure.
Kang et. al~\cite{Kang:2019:IEEEIoT:contract} accomplishes the pairwise contribution qualification by introducing reputation. users apply a multi-weight subjective  logic model~\cite{Liu:11:FGCS:logic} to obtain reputation of each other. A participant gains higher reputation by providing more positive actions that are recorded in a blockchain for transparency. 
Besides the pairwise direct reputation by users, there are also indirect reputation 
designed in this model using the records of multiple federators.
Lyu et. al propose FPPDL~\cite{Lyu:2019:iacr:Towards} and demonstrate similarity-based qualification by differential privacy generative adversarial networks (DPGAN)~\cite{Lu:2017:CCS:DPGAN}. In FPPDL, data provider generates artificial samples with DPGAN, and data verifier uses its local model to implement cross-user labeling. Then, the verifier computes the contribution measure by the label similarity between the data provider and verifier.

\subsection{Incentive Mechanisms as Reaction}

Here, we introduce incentive mechanism that rewards and reacts to honest participants with different quality based on contribution measurement. Firstly, we rigorously define FL incentives to give a clear understanding. Then, the various goals of incentive design are provided. Moreover, we also survey game theory that is widely adopted incentive design.

\begin{figure}
\setlength{\abovecaptionskip}{-0cm}   
\includegraphics[height=7cm]{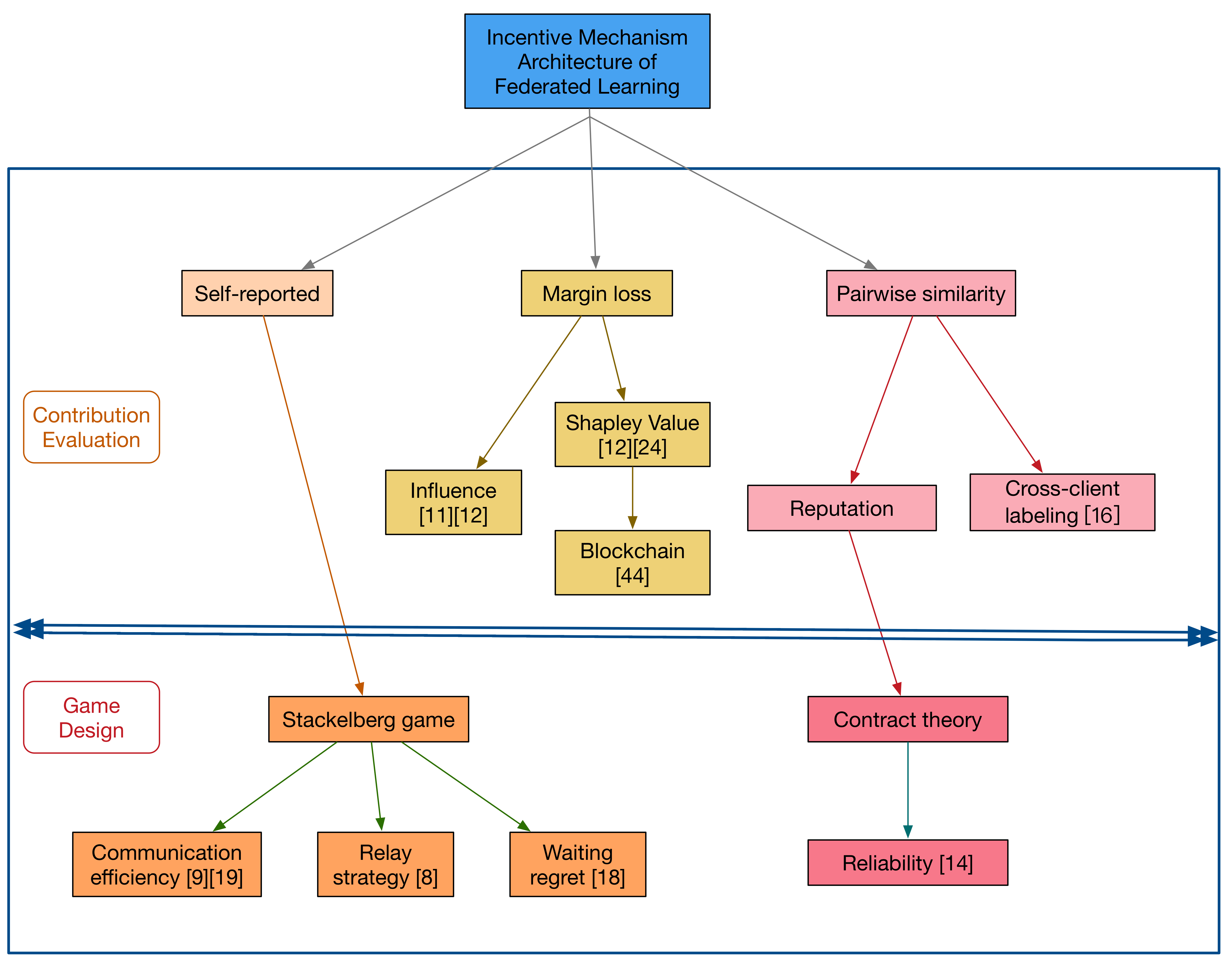}
\caption{Recent studies on incentive mechanism of federated learning.}
\label{incentive}
\end{figure}

\subsubsection{\textbf{Definition of Incentive Mechanism in FL}}

Incorporating the ideas from a multitude of studies on incentive mechanism of federated learning, we propose the first formal definition of incentives for FL. 


\textit{Incentive Mechanism of Federated Learning:} An incentive mechanism in FL system consists of a set of rewards $R$ and three functions $v$, $c$, and $r$. The function $v \colon R \rightarrow \mathbb{R}$ assigns each reward a value.  
For a set of participants $P$, the function $\contribution: P \rightarrow \mathbb{R}$ assigns each participant a score that measures their contribution to the system. We discussed different contribution measurement strategies in the previous subsection. 
Last, the function $r: \mathbb{R} \rightarrow R$ assigns a reward based on the score that $c$ provides. The reward function $r$ offers rewards of monotonously increasing value, i.e., if $x>y$, then $(r(x))\geq (r(y))$. 

From this definition, we see that the incentive design of federated learning include two main procedure: 1. \textit{Contribution Measurement}, which is discussed above;
and 2. \textit{Rewards (punishments) Allocation}. The FL systems deliver rewards based on the contribution using profit distribution methods including game theory and blockchain. Rewards could be monetary reward, generally, or other schemes such as biased information.

\subsubsection{\textbf{Goals of Designing Incentive Schemes}}

Based on the definition above, we examine that the incentive mechanism designs for federated learning attempt to encourage desirable behavior in users. More specifically, goals of an incentive scheme generally include two main factors:

\textbf{Attract Users of High Local Quality.}
The aggregated results of federated learning highly depend on the quality of participated users, including local data size and computation resources. An incentive scheme should attract users of high quality to join, such that the global model achieves good performance. On the other hand, data owners of low quality are supposed to be discouraged from joining due to the low revenue the incentive mechanism offers. 
    
\textbf{Attract Users with Good Networking Resources.}
Network transmission condition of users and the federator, or between users are also supposed to be taken into consideration while designing incentive mechanism, since both effectiveness and efficiency are imperative for the system performance. Additionally, some systems apply incentive mechanism to enhance some specific characteristics according to the objectives of these distributed systems. For instance, \cite{yu:2020:AIES:fair} solves the issues of costs and temporary mismatch between contributions and rewards to model users' regret user. Other examples (\cite{Khan:2019:corr:edge} ,\cite{Pandey:twc:2020:OnDevice}) focus on improving the communication efficiency of
    federated learning systems through involving transmission time as a highly weighted  factor in the utility function of incentive mechanism.


\subsubsection{\textbf{Incentive Design}}
 When participating in FL, users aim to maximize their rewards through incentives in comparison to the data and resources they provide to the system. 
 Given their specific local situation, each participant hence has a utility function they aim to maximize. 
 In order to determine the best way of action, participants consider possible action plans for themselves and the other participants. From a network resource perspective, the overall goal is to maximize collective utility.
 As a consequence, game theory is a useful methodology to design and analyze FL incentives. In the following, we discuss the different assumptions about participants and their relations in the context of the resulting games and incentives.

\textbf{Stackelberg Equilibrium in Non-cooperative Game.}
Stackelberg games are of use if one of the players is in a leader position while the others are followers. Thus, they are quite suitable for FL as the federator can be seen as the leading party. 
In a Stackelberg game, the followers usually first observe the behavior of the leading party before deciding their own actions.
Concretely, the leader decides an output, and then the followers can observe this to determine their own output factors such as resource inputs.
A limitation of the game is the assumption that the leader should be able to fully apprehend the behavior of the followers and thus needs to be aware of their local utility functions. 
Thus, the output determined by the leader is a profit maximization constrained by the utility function of the followers. In this strategy, the non-cooperative framework assumes all participants act separately. 
.

As the data of the participants in FL is not available to the federator, the federator does not know the utility functions with regard to data. Thus, Stackelberg games are only relevant when incentivizing the contribution of network resources. 
So, they can be applied to mitigate
the delays in completion of each training batch
by analytically obtaining equilibrium solution of a Stackelberg game\cite{sarikaya2019motivating}.

Another Stackelberg game-based approach \cite{Pandey:twc:2020:OnDevice} handles the communication efficiency of users implementing an uncoordinated computation
strategy during model aggregation. Specifically, it models a two-stage Stackelberg game by establishing a communication-efficient cost model for users and a reward rate for the federator.


Resources are particularly important in the context of edge and IoT due to the restricted capacity of the devices. Here, Stackelberg games have been suggested for user utility functions depending on the number of local iterations, i.e., local computation power \cite{Khan:2019:corr:edge}. 
In contrast, the federator aims at maximizing its utility in terms of the global model, trying to, e.g., minimize the number of communication rounds needed to reach a desirable global accuracy. However, there is not any concrete utility function in this work.

Other studies focus on very specific scenarios for FL. In the absence of direct communication between all participants, incentives for adapting a relay network can be modeled as Stackelberg games \cite{feng:2019:ithings:relay}. 
However, in a
cooperative relay network design, a larger training data set can result in a lower probability to be relayed due to its higher bandwidth use. As a result, the learning service pricing and cooperative relaying should be considered jointly.
Moreover,  \cite{yu:2020:AIES:fair} makes the assumption that the rewards can only be paid once the federation has made a gain from their model. It thus studies the payoff-sharing scheme on costs and temporary mismatch between contributions and rewards of FL, focusing on waiting time fairness.
Their proposed scheme FLI maximizes the overall effectiveness of the data alliance, and at the same time minimizes the imbalance of regret between users of delays caused by the training and commercialization time.

\textbf{Contract Theory Application.}
Contract theory is an economical theory that regards all transactions and institutions as a kind of contract.
It then designs the optimal contract to reduce the moral hazard, adverse selection, and extortion of the parties under the condition of asymmetric information, so as to ultimately improve social welfare. 
Contract theory can either deal with complete contracts \cite{Holmstrom:2012:complete}, meaning that 
the predefined contract specifies the legal consequences of every possible state, or incomplete contracts, which includes consideration of the incentive effects of parties' inability to make complete contingent contracts \cite{Hart:1988:Econometrica:Incomplete}.

In federated learning systems, complete contract theory~\cite{holmstrom1991multitask} has been applied due to its clear decision tree of responsibilities and obligations. The federator determines the contract items and users choose appropriate contract types based on their own resources to maximize profits on each side. 
Thus, contract theory is a type of Unbalanced Stackelberg game, with the federator as the leader and dominant
the optimization objective of the federated learning system.
However, the federator provides multiple optional contract classes for contract theory-based incentives, which is not possible using Stackelberg games to enhance rewarding efficiency.

Incentive schemes based on contract theory are more robust than Stackelberg game, in terms of computational complexity.
They allow to simulate data market transactions more realistically and avoid some unnecessary fine-grained operations to enhance efficiency of the federated learning systems.
Concretely, contract theory allows the user to select the function that maximizes its own utility based on the evaluation of the quantity, quality, computing resources, and communication capacity of the local data. To maximize its global profit, the federator takes the computation and communication efficiency and model accuracy of the uploaded gradient by users into account. However, verifying the authenticity and quality of the uploaded updates provided by the users remains difficult.


For incentive studies based on contract theory in federated learning, Kang et al~\cite{Kang:2019:IEEEIoT:contract} address the challenges of incentive mechanisms
for participating in training and worker selection schemes for reliable federated learning. It introduces reputation as the metric to measure the reliability and trustworthiness of the mobile devices and combine contract theory to motivate high-reputation mobile devices with high-quality data.

\textbf{Shapley Value in Cooperative Game.}
The above games are from the perspective of the federator and are based on leadership competition or non-cooperative games. 
An alternative approach is given by cooperative games: the profit of at least one party increases without reducing the profits of other parties.
Thus, the total utility increases with the participating of multiple members.
The key methodology here is the Shapley Value~\cite{Roth:1988:Shapley}, which evaluates the contribution of a participant as loss experienced by the participant leaving. In this manner, the Shapley value is independent of the order in which participants join. It assigns a unique distribution among the parties of a total surplus generated by the coalition of all members.
Furthermore, the Shapley Value allows using a combination of desirable properties to define a participant's contribution rather than focusing on one property. 


Formally, we denote the data federation $F=<Users, v>$ has been contributed by several users as $Users=\{U_1, U_2, ...,U_i\}$,
where $v$ is the contribution value function of this system. In federated learning scenarios, it could be the aggregated model accuracy. The Shapley Value define the contribution of $U_i$ to join $Users$ in $F$ as a margin loss despite the joining sequence as: 

\begin{equation}
    \delta U_i(users) = v\{users \mathop{\cup} U_i \} -v\{users\}
\end{equation}

Since the Shapley Value makes a fair distribution regardless of the joining order, there are $|Users|!$
joining sequences with corresponding probabilities. The probabilities of each sequence (or coalition) $S$
containing $User_i$ could easily be obtained by $|S|!(|F|-|S|-1)!/|F|!$ 
Thus, the contribution of $C_i$ by Shapley Value is:

\begin{equation}
        SV(F, C_i) = \sum_{S \subseteq F \setminus \{C_i\}} \frac{|S|!(|F|-|S|-1)!}{|F|!}\delta C_i(F)
        \label{sv}
\end{equation}

There are a number of studies that use Shapley Value for their incentive design. 
In vertical FL, it has been used to calculate the grouped feature importance since features are grouped to join data federation by multiple users \cite{wang:2019:Bigdata:contribution}.
Although Shapley Value also works for horizontal FL, the reason why the authors apply Influence function is that we need to note is that Shapley Value based distribution solution often takes exponential time to compute with a huge complexity of $O(n!)$, where $n$ denotes  the user size.
Nevertheless, this method also sheds light on the researches in model contribution using Shapley Value in the context of federated learning.

The key challenge of computing the Shapley Value lies on the need for extra training to compute the marginal contribution of a user. A contribution index that reconstructs the approximate models on different combinations of the datasets through the intermediate results during the training process replaces the exact Shapley value \cite{Song2019Profit}. In this manner, efficient contribution measurement becomes possible. 

Last, Shapley Value has been used in combination with a blockchain network due to its fairness and high computation overhead. The party who can decide on a new block is selected based on their Shapley Value\cite{wang:2019:Bigdata:contribution}.



\section{Malicious User Updates: How to Detect and Limit the Damage}
\label{sec:MaliciouscontribQuality}
In FL frameworks, machine learning tasks are massively distributed among participants. 
Ideally, this large-scale distribution helps ML-service providers reach more diversified data sources and thus build stronger models. 
Nonetheless, in the basic design principals of Federated Learning, user selection is mainly based on users' data availability, their computational power, and network resources, without any solid guarantees on user reliability or trustworthiness~\cite{bonawitz2019towards}. 
As a consequence, Federated Learning can be subject to various client-side attacks with different objectives.\\
It has been shown in prior art~\cite{} that participants might deviate from the intended FL protocol and try to bring damage to the ecosystem. This malicious activity varies from simple selfish user behavior to intentionally sending faulty contributions to tamper with the federated model.\\
In the following, we characterize types of malicious user contributions that might intentionally deteriorate model quality and survey existing detection and prevention mechanisms that protect against them.

\subsection{Malicious Behaviour Characterization Criteria}
Multiple state-of-the-art works have been proposed to demonstrate the damage caused by malicious
participants in Federated Learning. It is worth mentioning that the attacks discussed in this section are carried out during training time either by insider malicious participants or by outsider adversaries that take over honest participants' devices.
Threats are characterized according to the following criteria. 

\textbf{Adversarial Goal.}
Participants can maliciously contribute to FL frameworks for a myriad of goals ranging from provoking arbitrary damage to the system to targeted causative violations. Offenders might try to prevent model convergence, deteriorate model accuracy, incorporate backdoors in the model, miss-classify a certain type of inputs, or even have access to the model without actually participating in the training process.

\textbf{Number of Offenders.}
Adversarial behavior can be carried by individual participants separately or multiple participants simultaneously. The latter can either be controlled by the same malicious party in order to bring more damage to the system (Sybil Attacks) or can collude to achieve a common adversarial goal.
  
\textbf{Participants’ Background Knowledge.}
The background knowledge of the attacker is a deterministic factor of the attack severeness. For instance, they may know other honest participants' training data or their training parameters. They can be aware of the mechanism applied by the federator to detect malicious activity or of the global data distribution, and so on.

\textbf{Attack Duration.}
Some FL malicious behavior may require to be carried out continuously through multiple rounds to take effect. In this case, the attack is said to be stealthy. On the other hand, some adversarial goals are more straightforward to achieve and thus the attack can be carried out in a single round.

\subsection{Characterizing Malicious Behaviour in FL}
\label{ssec:MaliciousBehavior}

\begin{table*}
\centering
\resizebox{\linewidth}{!}{%
\begin{tabular}{|c|c|c|c|c|c|} 
\hline
\begin{tabular}[c]{@{}c@{}}Attack Category\end{tabular}                               & Attack                                            & \begin{tabular}[c]{@{}c@{}}Adversarial Goal\end{tabular}                                                                                                                                                  & \begin{tabular}[c]{@{}c@{}}Number of\\Offenders\end{tabular} & \begin{tabular}[c]{@{}c@{}}~Participants’ Background\\Knowledge\end{tabular}                                                                                           & \begin{tabular}[c]{@{}c@{}}Offense\\Duration\end{tabular}  \\ 
\hline
\multirow{6}{*}{\begin{tabular}[c]{@{}c@{}}Targeted Poisoning\\Attacks\end{tabular}}   & \cite{ bhagoji2019analyzing}     & \begin{tabular}[c]{@{}c@{}}Provoke targeted misclassification~\\and negate the combined~effect of benign agents~\end{tabular}                                                                              & \begin{tabular}[c]{@{}c@{}}Single~\\attacker\end{tabular}    & \begin{tabular}[c]{@{}c@{}}White-box access to the model,\\Access to training~data~\end{tabular}                                                                        & Stealthy                                                   \\ 
\cline{2-6}
                                                                                       & \cite{ liu2020backdoor }         & \begin{tabular}[c]{@{}c@{}}Assign an attacker-chosen label\\to input data with a specific trigger~\end{tabular}                                                                                            & \begin{tabular}[c]{@{}c@{}}Sybil\\attack\end{tabular}        & \begin{tabular}[c]{@{}c@{}}White-box access to~the model,\\Access to training data,\\Access to a portion of a subset\\of the feature space\end{tabular}                 & Stealthy                                                   \\ 
\cline{2-6}
                                                                                       & \cite{ chen2020backdoor}        & \begin{tabular}[c]{@{}c@{}}Introduce a persistent change in a\\joint meta-learning model such that,\\when a user adapts it for a new classification\\task, targeted misclassification occurs~\end{tabular} & \begin{tabular}[c]{@{}c@{}}~Single\\attacker\end{tabular}    & \begin{tabular}[c]{@{}c@{}}~White-box access~to the model,\\Access to training~data~~\end{tabular}                                                                      & One-shot                                                \\ 
\cline{2-6}
                                                                                       & \cite{ tolpegin2020data}         & \begin{tabular}[c]{@{}c@{}}Provoke high testing errors for\\particular subset of classes\end{tabular}                                                                                                      & \begin{tabular}[c]{@{}c@{}}Sybil\\attack\end{tabular}        & \begin{tabular}[c]{@{}c@{}}White-box access\\to the model,\\Access to training~data\end{tabular}                                                                        & Stealthy                                                   \\ 
\cline{2-6}
                                                                                       & \cite{ bagdasaryan2020backdoor } & \begin{tabular}[c]{@{}c@{}}Inject a backdoor task in the\\model\end{tabular}                                                                                                                               & \begin{tabular}[c]{@{}c@{}}Single\\attacker\end{tabular}     & \begin{tabular}[c]{@{}c@{}}White-box access\\to the model,\\Access to training~Data,\\Knowledge regarding the \\detection mechanism used\\by the federator\end{tabular} & One-shot                                                   \\ 
\cline{2-6}
                                                                                       & ﻿\cite{cao2019understanding}     & \begin{tabular}[c]{@{}c@{}}Provoke high testing errors for\\particular subset of classes\end{tabular}                                                                                                      & \begin{tabular}[c]{@{}c@{}}Sybil\\attack~~\end{tabular}      & \begin{tabular}[c]{@{}c@{}}White-box access\\to the model,\\\`{}Access to training~Data\end{tabular}                                                                    & Stealthy                                                   \\ 
\hline
\multirow{2}{*}{\begin{tabular}[c]{@{}c@{}}Untargeted Poisoning\\Attacks\end{tabular}} & \cite{fang2019local}            & \begin{tabular}[c]{@{}c@{}}Cause a high miss-classification\\rate~\end{tabular}                                                                                                                            & \begin{tabular}[c]{@{}c@{}}Sybil\\attack\end{tabular}        & \begin{tabular}[c]{@{}c@{}}White-box access\\to the model,\\Access to training~Data,\end{tabular}                                                                       & Stealthy                                                   \\ 
\cline{2-6}
                                                                                       & ~\cite{li2020learning}~          & \begin{tabular}[c]{@{}c@{}}Degrade the overall model\\performance~\end{tabular}                                                                                                                            & \begin{tabular}[c]{@{}c@{}}Sybil\\attack~\end{tabular}       & White-box accessto the model                                                                                                                                            & Stealthy                                                   \\ 
\hline
\begin{tabular}[c]{@{}c@{}}Free-Rider \\Attacks\end{tabular}                           & \cite{lin2019free}              & \begin{tabular}[c]{@{}c@{}}Have access to the model without\\participating in the training\end{tabular}                                                                                                    & \begin{tabular}[c]{@{}c@{}}Single\\attacker\end{tabular}     & \begin{tabular}[c]{@{}c@{}}White-box access to the model,\\Knowledge of how normal\\~updates look like\end{tabular}                                                     & Stealthy                                                   \\
\hline
\end{tabular}
}
\caption{Characterization of malicious behavior in Federated Learning }
\label{tab:classificationAttack}
\end{table*}

In the following, we characterize three possible malicious participant contributions (summarized in Table~\ref{tab:classificationAttack}), that might negatively impact model quality in the FL ecosystem.  
We describe these attack categories according to the criteria defined above and survey the existing state-of-the-art works that study them.

     \textit{\textbf{Targeted Poisoning.}}  In this type of malicious behaviour, an attacker tries to inject a backdoor task of his interest in the global model along with the main task that was initially trained
     without deteriorating the model's accuracy.
     This adversarial goal can be achieved in two possible ways. The first one is generating poisonous data locally, carrying out local training on the malicious participant side using this faulty data, and then sending the resulting poisonous updates to the federator for aggregation. Generating poisonous data can be done by simply flipping labels  or by injecting naturally occurring or artificial patterns in the feature space that is associated with the backdoor. This malicious behavior is referred to as data poisoning attacks in the state-of-the-art ~\cite{funglimitations,tolpegin2020data,cao2019understanding,liu2020backdoor}. The second way is model poisoning where the attackers carefully craft poisonous updates that efficiently inject the backdoor task in the model~\cite{bagdasaryan2020backdoor,fang2019local,chen2020backdoor,bhagoji2019analyzing}. Both of these attacks can be done by a single participant individually or by multiple sibyls collaboratively~\cite{sun2020data,li2020learning,tolpegin2020data,funglimitations,cao2019understanding,bagdasaryan2020backdoor}. To achieve model poisoning, malicious participants might send faulty contributions over multiple training rounds till the damage is done while the most severe attacks can successfully inject the backdoor in a single round \cite{bagdasaryan2020backdoor}.
     
   \textit{\textbf{Untargeted Poisoning.}} Unlike targeted poisoning, in this category of malicious behaviour, the attacker's goal is to cause a high miss-classification rate indiscriminately for testing samples. As a consequence, the learned model is unusable and hence the attack is essentially a denial-of-service attack. Generally, the malicious participant does not need to carry out data poisoning but can simply craft model updates that provoke severe accuracy drop. Concrete instantiations of this type of attack in the federated learning setting include \cite{fang2019local,li2020learning}. The impact on model accuracy can be even more aggressive when the attacker is aware of the detection mechanism used on the federator's side \cite{fang2019local} since it can adapt the pace of sending malicious contributions to remain undetected (up to 78\% accuracy drop \cite{fang2019local}).
   
   \textit{\textbf{Free-rider.}} In this category of malicious behaviour, self-interested participants want to take advantage of the federated learning service without actually participating in it due to the lack of data, lack of computing resources, or even for privacy concerns. To do that, free-riding participants craft fake updates via simple random generation or based on previous versions of the model to pretend that they participate in the learning process. Even though this kind of behavior has been widely explored in the case of peer-to-peer systems, there is only one state-of-the-art work that explores how it applies to federated learning \cite{lin2019free}.
   Although the presence of free-riders in FL-based frameworks might seem harmless, the behaviour of this category of participants is opposite to the main purpose of federated learning which consists of doing large scale distribution of ML-based tasks to have access to more diversified and rich data sources. Free-riders can either have no novel contributions to the system or in worse scenarios send arbitrary updates that might negatively impact the trained model's accuracy.


\subsection{Defense Mechanisms Against Malicious Contributions}
\label{ssec:defence}
There are two possible ways to protect against malicious contributions in Federated Learning. On one hand, the federator can implement detection mechanisms and punish attackers once he suspects an anomaly. He can either react by reducing their learning rate gradually or directly evict them from the system. On the other hand, the basic Federated Learning protocol can be enhanced by prevention mechanisms that stop malicious behavior from occurring in the first place. We present below some state-of-the-art mechanisms that were proposed to detect and prevent malicious contributions in FL frameworks.


    \textit{\textbf{Gradient Auditing.}} The purpose of this kind of protection mechanism is to detect and punish malicious behaviour such as model poisoning or free-riding. In this case, the federator is assumed to be trusted and he monitors statistical changes in model updates. The latter tries to point out suspicious updates, and exclude them from the aggregation process or reduce their weights. Examples of such approaches are FoolsGold \cite{sun2019can} and Gradient Norm Bounding \cite{fung2018mitigating}. 
  
    \textit{\textbf{Trusted Execution Environments.}} This a hardware-based protection mechanism that is mostly adapted to cross-silo \footnote{Cross-silo Federated Learning is an FL setting that involves a small number of relatively reliable clients, for example multiple organizations collaborating to train a model.} federated learning ecosystems where the local training code on the participants-side is implemented in a Trusted Execution Environment (TEE) such as Intel-SGX (e.g., \cite{lie2017glimmers}). This way, the code run by participants is certified by the federator to make sure that the updates they send are not malicious. Thus, trusted execution environments prevent any attempt at deviation from the intended FL protocol. 
    
    \textit{\textbf{Gradient Sparsification.}}
    This protection mechanism limits the effect of causative attacks in federated learning by pruning gradients that have small magnitude, this is also referred to as gradient compression. It has been shown in \cite{lin2017deep}  that gradients can be compressed up to a factor of 300, while maintaining the same model accuracy. This approach was initially proposed to reduce communication bandwidth in distributed learning but was proved in \cite{liu2020backdoor} to be an effective way to protect against targeted poisoning with a reasonable accuracy-loss/protection-level tradeoff.
    
    \textit{\textbf{Differential Privacy}}
    Initially, differentially-private FL was proposed to reduce information leakage about local users’ data \cite{geyer2017differentially}. 
    However, since adding noise to user updates bounds their influence over the joint model, some state-of-the-art works \cite{bagdasaryan2020backdoor, liu2020backdoor} considered using differential privacy as a protection mechanism to limit the damage caused by poisoning attacks.
    This approach works by first clipping amplified and potentially malicious updates, then adding Gaussian or Laplacian noise to them.
    This simply reduces the impact of causative attacks but does not entirely eliminate them. Also, adding user-level noise potentially reduces the accuracy of the trained models.


\section{Empirical Analysis}
\label{sec:contribQuality}
Here, we aim to quantify how the existing contribution measurement strategies could recognize attackers and their stability under attackers. Specifically, we consider a scenario of federated training image classifier with benign and malicious users. We implement three popular strategies against the attack of flipping labels.
   
\vspace{-0.3cm}
\subsection{Experiment Setup}


 The Federated Learning system under evaluation consists of one federator and 4 users. 
The model to be trained is a VGG-type\cite{Simonyan15} convolutional neural network (CNN).
Each user possesses 6000 unique data samples randomly selected from the CIFAR10 dataset~\cite{Krizhevsky09learningmultiple}. Original CIFAR-10 dataset consists of 60000 32x32 colour images in 10 classes, with 6000 images per class.

Some of the users are malicious and perform a data poisoning attack. 
When the attackers train their local models, they inject data noise by flipping the label with a probability $p$. 
The label is flipped with one of the other 9 labels randomly. 

The flipping probability $p$ is varied between  
10\%, 30\%, 50\% and 100\%. 
The number of attackers is varied between 0 and 3. 

In the following, we evaluate the user data contribution to the global model with three mechanisms:

\begin{itemize}
    \item \textbf{Influence}: The classic notion of Influence means to measure the effects on global accuracy of individual data points \cite{cook1980characterizations}. 
    Denote the global aggregated model as $\hat{\theta}$ and the global model $\hat{\theta} _{/i}$ without the user $U_i$
    as $\hat{\theta} _{/i}$. 
    The contribution for a data set $T$ is then quantified as the difference in accuracy between the two models, i.e., 
   $inf(U_i ,T,\theta) = Acc({T,\hat{\theta}})-Acc({T,\hat{\theta}}_{/i})$. 
    
    \item \textbf{Reputation}: Similar to Influence, Reputation quantifies the influence of each user.
    However, the score assigned is binary with 1 indicating that the involvement of user $C=U_i$ improves global accuracy. Reputation considers several time slots (similar to global rounds).
    In our experiment, there are $ts=5$ time slots and we average the contribution measurement of user $U$ as
    $Rep(U_i ,T,\theta) = \frac{1}{ts} \sum_{ts}H(Acc({T,\hat{\theta}})-Acc({T,\hat{\theta}_{/i}}))$, where H(x) is the heaviside unit step function.

    \item \textbf{Shapley}: In the settings of Shapley Value, we follow the definition and calculation of Equation \ref{sv}. Four users join this training process and the federator determines their contributions by sequential deletion of marginal loss. The Shapley Value could see the impact on joining order of different users in federated learning. 
\end{itemize}

The reason why we present the evaluation details of Shapley in Section \ref{incentive} while the others above is that Influence and Reputation are relatively straight forward and we just need to specify some parameters. However, Shapley evaluation is also a solution to Cooperative Game whose algorithm is well defined in existing studies. Note that all three mechanisms are marginal loss-based, as the other types of approaches like self-reporting are obviously unable to deal with attacks. 

The experiments are conducted with library Keras-2.3 based on Tensorflow-2.2, and executed on Dell Alienware Aurona (20 CPUs with 32G RAM) equipped with one RTX 2018 Ti GPU.

\vspace{-0.3cm}
\subsection{Experimental Evaluation}









\begin{figure}[htb]
	\centering
	\includegraphics[width=\linewidth]{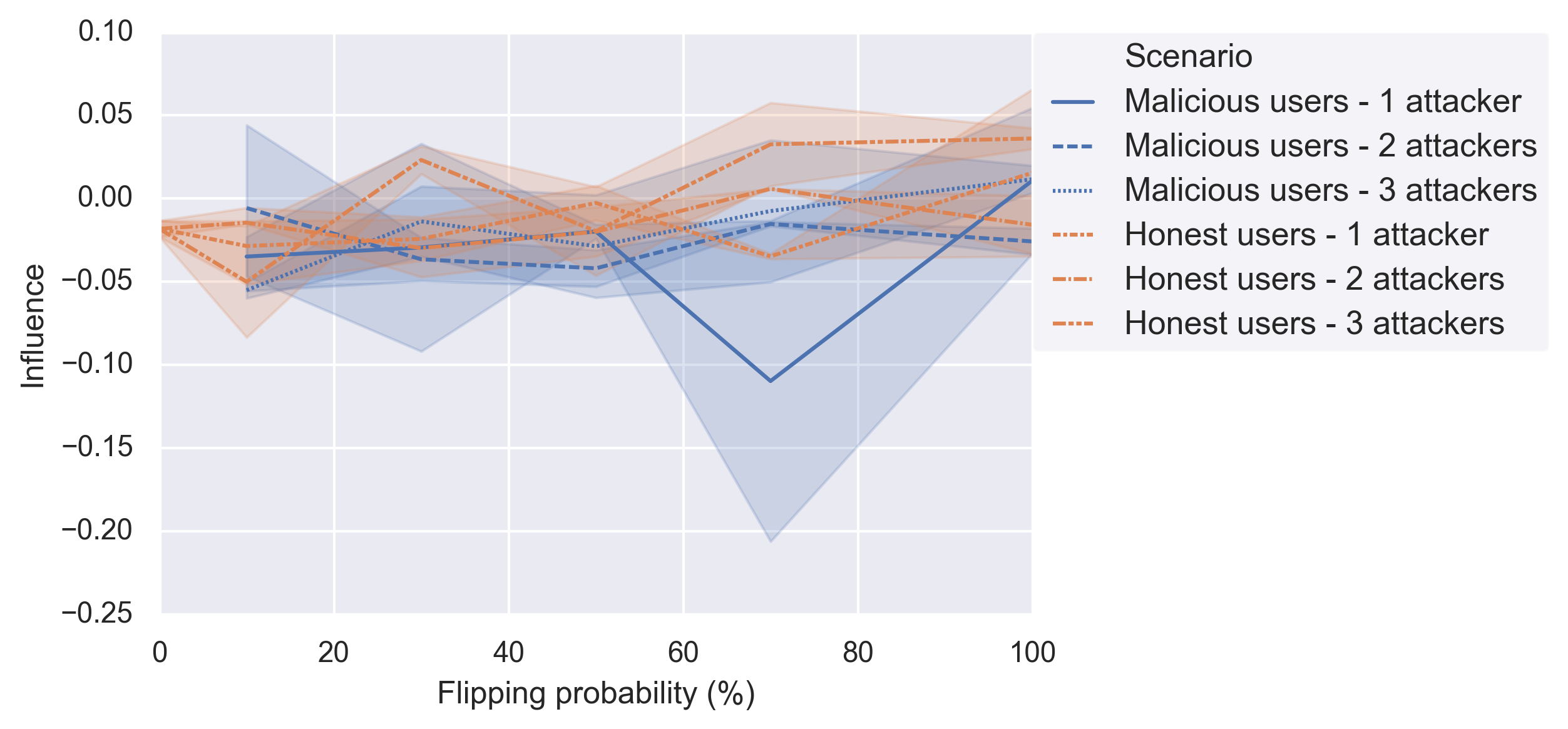}
	\caption{ Influence mechanism under combinations of users and attackers. A higher value indicates a higher contribution.
	}
	\label{fig:Influence}
\end{figure}
\begin{figure}[htb]
	\centering
	\includegraphics[width=\linewidth]{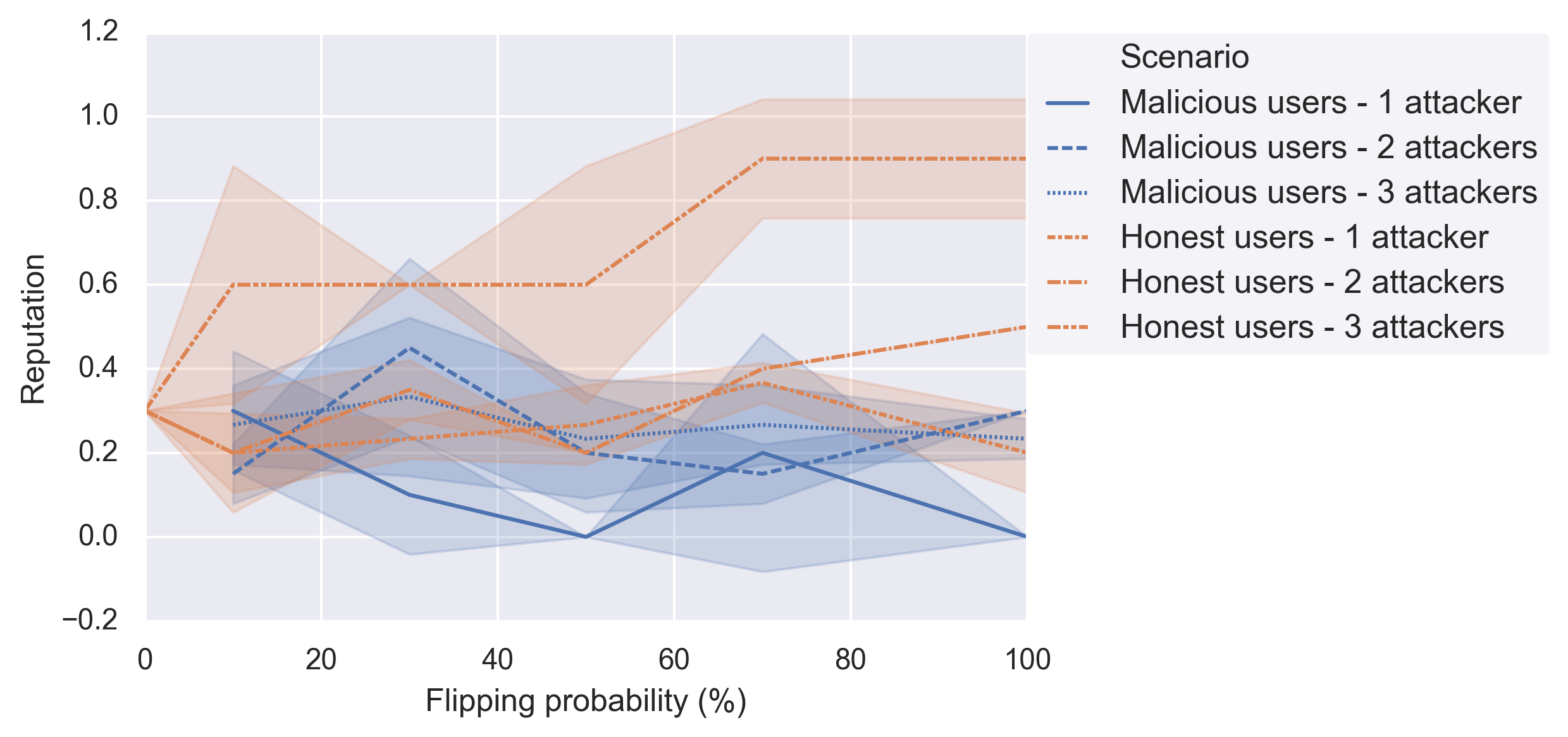}
	\caption{Reputation mechanism under combinations of users and attackers. Higher the value, better the reputation.}
	\label{fig:Reputation}
\end{figure}
\begin{figure}[htb]
	\centering
	\includegraphics[width=\linewidth]{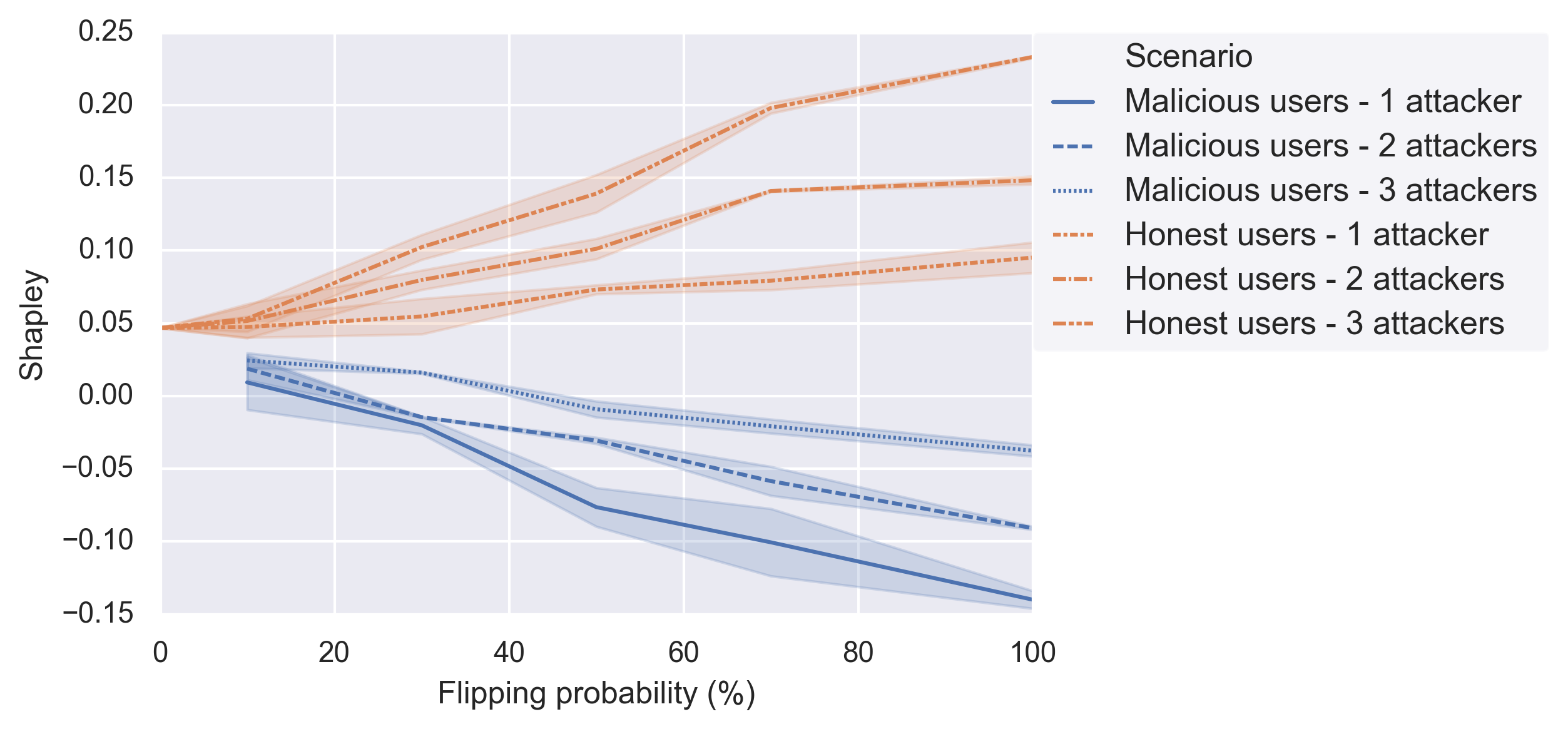}
	\caption{Average Shapley Value under combinations of users and attackers. Higher the value, better the contribution. }
	\label{fig:Shapley}
\end{figure}

Figure \ref{fig:Influence} - \ref{fig:Shapley}
display the measured user's data contribution with respectively Influence, Reputation, and Shapley.
In our experiments, we vary the number of attackers, and we vary the flipping probability $p$. Here, the values of Reputation have been normalized
into $[0,1]$.
Thus, generally, all three strategies succeed in recognizing attackers since we could see from Figure \ref{fig:Influence} - \ref{fig:Shapley} that the mean values of averaging honest users are larger than those of attackers. It demonstrates the effectiveness of contribution measurement approaches based on marginal loss and similarity. Additionally, overall results also show that for malicious users, higher flipping rates may result in lower measured contribution, which is more stable in Figure \ref{fig:Shapley} while there are fluctuations in Figure \ref{fig:Influence} and \ref{fig:Reputation}. And if we consider a given flipping rate, e.g., 3 attackers with a flipping rate of 50\%, the Influence in Figure \ref{fig:Influence} of honest users and the Influence of malicious users are almost the same with 3 attackers with a flipping rate of 10\%. This exhibits the fact that such techniques (similar in Figure \ref{fig:Reputation}) to quantify user contribution is not pertinent in the case of malicious users.

Comparing the three figures, Shapley measurement in Figure \ref{fig:Shapley} shares the highest capability while Influence in Figure \ref{fig:Influence} finds difficulty
in recognizing attackers. This is reasonable since Reputation in Figure \ref{fig:Reputation} qualifies and sums up influence values in multiple rounds, which also indicates the potency after multiple global iterations of both strategies. We could also observe that especially in Figure \ref{fig:Shapley} and Figure \ref{fig:Reputation}, the average value on honest and malicious users share opposite trends on the value with increasing flipping ratio. The diversity indicates the implicit relativity between the contribution of the honest and the malicious users since they are all based on marginal loss.
As for Shapley, the significant difference to
Influence illustrates the importance of the impact of joining sequences in federated learning.
In addition, similarly, the 
variety on different flipping level shows more discrepancies and conforms most to our theoretical prospective
on Shapely than Influence and Reputation. 


\vspace{-0.2cm}
\section{Research Directions}
\label{sec:researchDirections}
We have seen that incentives in federated learning require consideration of malicious behavior as they are not necessarily able to detect such behavior. In this section, we outline research directions to investigate this research gap which we believe are promising.
\vspace{-0.3cm}
\subsection{Novel Attack-Aware Incentives}
 As indicated by the results in Section~\ref{sec:contribQuality}, designing new incentive mechanisms should consider attacks. One possible solution may be introducing blockchain-based contribution measurements with transmitted parameter records on chain~\cite{Yuan:20:fedcoin}.
 Indeed, such a incentive mechanism can possibly be used to detect attackers as those users  achieve low scores in the contribution measurement. After attack detection, malicious users can be evicted from the system to prevent future harm. 

\vspace{-0.3cm}
\subsection{Alternative Contribution Measurements and  Alternative Attacks}

Our experimental evaluation in this paper considered merely label flipping attacks and three contribution measurement approaches. 
Future studies should extend these results to other attacks and contribution measurement mechanisms. 
In Section~\ref{ssec:MaliciousBehavior}, we already identified untargeted poisoning and Free-riding attacks
as potential threats that require further consideration in the context of incentives. 
An example for a future study related to Free-riding is to evaluate
whether cross-user labeling recognizes attackers whose adversarial goal is to have access to the model without participating in the training. 
In particular, as all users just transmit and verify generated data based on their own data, an attacker can generate new data based on other submissions to appear as if they contribute. 

We can also evaluate these contribution measurement strategies in the presence of other non-causative active attacks that aim at inferring sensitive information about participants' data such as class-representatives~\cite{melis2019exploiting}, data distribution~\cite{wang2019eavesdrop}, etc. Although these attacks do not specifically target model quality, they may indirectly have an influence on it.

\vspace{-0.3cm}
\subsection{New Attacks Targeting Incentives}
In this paper, we primarily focused on the impact of attacks on model accuracy. 
Yet, Free-riding does not primarily target model accuracy but rather deals with parties that gain something without contributing appropriately. 
As stated in Section~\ref{sec:MaliciouscontribQuality}, Free-riding attacks are not yet fully explored in the context of Federated Learning. 
The work presented in~\cite{lin2019free} considers an adversarial model where lazy participants aim at using the federated model without actually being engaged in the training process. 
We believe that it could be interesting to explore other adversarial strategies for this attack category in the presence of incentive mechanisms. In the context of incentives, adversaries want to maximize the profit they gain out of the deployed incentive mechanism and simultaneously minimize the computational effort they have to invest into gaining from the mechanism. Concretely, a self-interested participant carefully crafts model updates that seemingly have high quality without doing actual local training.
This is a contrasting view of attack-aware incentive design in terms of adversarial goals that are equally undesirable as participants are less likely to be incentivized to contribute honestly if incentives can be gamed. 




\vspace{-0.3cm}
\section{Conclusion}
\label{sec:Conclusion}
Motivated by the increasing threat of malicious users on federated learning systems, we presented exploratory analysis on how contribution measurement strategy of incentive mechanisms can characterize attackers. We surveyed existing attacks on model accuracy and highlighted that they can have a detrimental impact on incentive measures. 
Through federatedly training a deep image classifier, we evaluated how simple label flipping attacks can degrade the performance of the state-of-the-art incentive measures.
Based on empirical evaluation and observations, we discuss future research directions. Specifically, it is imperative to design new incentive mechanisms that are resilient to novel attacks  circumventing the detection of incorrect data. We also highlight how free-rider attacks with the goal of gaining unjustified rewards is a largely unexplored but critical threat. 

\vspace{-0.3cm}
\section{Acknowledgment}
This work has been partly funded by the Swiss National
Science Foundation NRP75 project 407540\_167266.

\bibliographystyle{IEEEtran}
\bibliography{biblio}


\end{document}